\documentclass[]{aa}

\usepackage{amsmath}
\usepackage{graphicx}
\usepackage{amssymb}
\usepackage{natbib}
\usepackage{bm}
\usepackage{ulem}
\bibpunct{(}{)}{;}{a}{}{,}

\begin{document}

\title{Influence of phase-diversity image reconstruction techniques \\ on circular polarization asymmetries}
\author{A. Asensio Ramos \and M. J. Mart\'{\i}nez Gonz\'alez \and E. Khomenko \and V. Mart\'{\i}nez Pillet}

\offprints{aasensio@iac.es}

\institute{Instituto de Astrof\'{\i}sica de Canarias, 38205, La Laguna, Tenerife, Spain; \email{aasensio@iac.es}
\and
Departamento de Astrof\'{\i}sica, Universidad de La Laguna, E-38205 La Laguna, Tenerife, Spain
}
\date{Received ; Accepted}
\titlerunning{Circular polarization asymmetries with image reconstruction techniques}
\authorrunning{Asensio Ramos et al.}
%
%
\abstract
{Full Stokes filter-polarimeters are key instruments for investigating the rapid evolution of
magnetic structures on the solar surface. To this end, the image quality is routinely improved using 
a-posteriori image reconstruction methods.}
{We analyze the robustness of circular polarization asymmetries to phase-diversity image reconstruction
techniques.}
{We use snapshots of magneto-hydrodynamical simulations carried out with different initial
conditions to synthesize spectra of the magnetically sensitive Fe \textsc{i} line at 5250.2 \AA. We degrade the synthetic profiles
spatially and spectrally to simulate observations with the IMaX full Stokes filter-polarimeter. 
We also simulate the focused/defocused
pairs of images used by the phase-diversity algorithm for reconstruction and the polarimetric
modulation scheme. We assume that standard optimization methods are able to 
infer the projection of the wavefront on the Zernike polynomials with 
10\% precision. We also consider the less favorable case of 25\% precision.
We obtain reconstructed monochromatic modulated images that are later
demodulated and compared with the original maps.}
{Although asymmetries are often difficult to define in the quiet Sun due to the complexity
of the Stokes $V$ profiles, we show how asymmetries are degraded with spatial and spectral smearing.
The results indicate that, although image reconstruction techniques reduce the spatial smearing,
they can modify the asymmetries of the profiles, mainly caused by the appearance of spatially-correlated 
noise.}
{}


\maketitle

\section{Introduction}
The observation of Stokes profiles in spectral lines are a valuable
tool to infer information about the thermodynamical and magnetic properties of 
the solar plasma. This information is encoded in the amplitude and the
shape of the Stokes profiles. Therefore, it is important to avoid
any effect that perturbs the shape because it can crucially 
modify the information encoded in the profile.

Of special relevance is the asymmetry of the circular polarization profile, 
which is known to be related to the correlation between velocity and magnetic
field gradients along the line-of-sight \citep[LOS;][]{illing75}. This effect has been
exploited to build atmospheric models with such gradients to explain asymmetries 
around magnetic flux concentrations \citep{solanki88,grossman_doerth88,sanchez_almeida89,solanki93}.
The situation is especially relevant in the weakly-magnetized zones of the quiet
Sun away from active regions, where Stokes $V$ profiles present a variety of
shapes with strongly asymmetric profiles 
\citep{sigwarth99,sanchez_almeida00,khomenko03,socas_pillet_lites04,marian08,viticchie_1_11,viticchie_2_11}.

Earth-based observations are always affected by the disturbing effect of the atmosphere.
As a consequence, the diffraction limit of the telescope is practically
never reached. It is often impossible to overcome the 1" limit if the 
observations are not accompanied by an adaptive optics system and powerful post-processing
methods. For this reason, \cite{khomenko_shelyag05} and \cite{shelyag07} analyzed the effect
of spatial smearing on the Stokes asymmetries observed in the pair of Fe \textsc{i}
lines at 630 nm and at 1565 nm. They concluded that asymmetries are heavily disturbed
by the lack of spatial resolution. They even discovered that it is possible to find
regions in which the field polarity is different in the two spectral regions \citep{sanchez_almeida03,rezaei07}, something
fully attributed to the lack of spatial resolution. The fundamental reason is that
the shape of Stokes $V$ profiles in weakly-magnetized regions are very complex and
they change in scales much smaller than the resolution element of the largest Earth-based telescopes
even in the hypothetical absence of atmosphere.

Even if the telescope is put in a balloon at 40 km height like Sunrise \citep{sunrise10}, 
there is some remaining atmosphere (albeit small) which, together with the intrinsic aberrations of the telescope and 
instruments, can modify the observations. For this reason, the IMaX instrument
onboard Sunrise was designed to use the phase-diversity post-facto reconstruction algorithm 
\citep{paxman92,santiago_vargas09}. In this paper, we focus on the interesting problem of
testing how the spatial and spectral degradation produced by the Sunrise/IMaX combination
affects circular polarization asymmetries and if the post-reconstruction algorithms can
help us extract reliable information (comparable to the unperturbed synthetic case) about 
them from degraded data. To this end, we 
use profiles synthesized on 3D models of the solar photosphere. We degrade them to simulate the
observational conditions of Sunrise/IMaX and reconstruct the images using the
phase-diversity algorithm.

\section{Data processing}

\subsection{Simulations and spectral synthesis}
The snapshots that we have used correspond to individual time
steps of a 3D magneto-hydrodynamical simulation of solar
magneto-convection done with the MURAM code
\citep{vogler_thesis03,vogler05,cameron_11}. An initially vertical
magnetic field of 200 G strength was introduced into already
developed purely hydrodynamical convection. The simulation box was
split into 4 parts with the opposite polarities of the magnetic
field in the adjacent parts. This magnetic field evolved self-consistently
with convective motions. The redistribution of the magnetic field led to
almost exponential decrease with time of its average unsigned
value over the simulation domain. The snapshots used in the
present work were taken 17, 36 and 112 minutes after the magnetic
field was introduced. At these time moments, the average unsigned
magnetic field strength at photospheric base was of 140 G, 80 G
and 30 G. The last snapshot is representative of quiet Sun
internetwork because it is close to the value obtained by
\cite{khomenko05}, while the first two are representative of bipolar
enhanced network. The computational box has $288 \times 288 \times
100$ grid points, with the largest number of points in the
horizontal directions. The grid step in the horizontal direction
is roughly 20 km, with a total horizontal extent of $6 \times 6$
Mm$^2$. The grid size in the vertical direction is $\sim$14 km,
with an ensuing total height of 1400 km. On average, the spatial
extent above and below optical depth unity at 5000 \AA\ amounts to
600 km and 800 km, respectively.

We have computed the Stokes spectra of the Fe \textsc{i} 5250.2089 \AA\
line formed at solar disc centre ($\mu=1$). The calculation was
done using the SIR code \citep{sir92} in LTE
approximation. The spectra were computed in a 2.5D approach, i.e.
individually for each vertical column of the snapshots,
corresponding to the LOS direction. We used the following atomic parameters: lower
level energy $E_l=0.121$ eV, $\log gf
=-4.96$ and iron abundance $A_{\rm Fe}$=7.50. 
The detailed description of this particular simulation run and
analysis of Stokes spectra of this and other Fe \textsc{i} lines can be
found in \citet{khomenko_shelyag05} and \citet{khomenko_collados07}.

Since we simulate the effect of image degradation on the polarimetric 
measurements, we first have to modulate the resulting monochromatic synthetic images 
of the Stokes parameters. These modulated images will be distorted
by the presence of aberrations and measured on the CCD. For simplicity, we apply 
the ideal modulation scheme:
\begin{equation}
\mathbf{O} = \frac{1}{\sqrt{3}}\left[
\begin{array}{rrrr}
1 & 1 & 1 & 1 \\
1 & 1 & -1 & -1 \\
1 & -1 & -1 & 1 \\
1 & -1 & 1 & -1
\end{array}
\right],
\end{equation}
and neglect any cross-talk between Stokes parameters.
Differences between the real modulation matrix for IMaX and the ideal one presented
above will fundamentally affect the efficiency of the modulation scheme. Consequently,
the only difference lies on the final noise level. The resulting modulated images are spatially degraded, noise is 
added and the reconstructed image is obtained through standard phase-diversity
formulae. The ensuing Stokes parameters are obtained by
applying the demodulation matrix $\mathbf{D}=\mathbf{O}^{-1}$.

\begin{figure}[!t]
\centering
\includegraphics[width=0.5\textwidth]{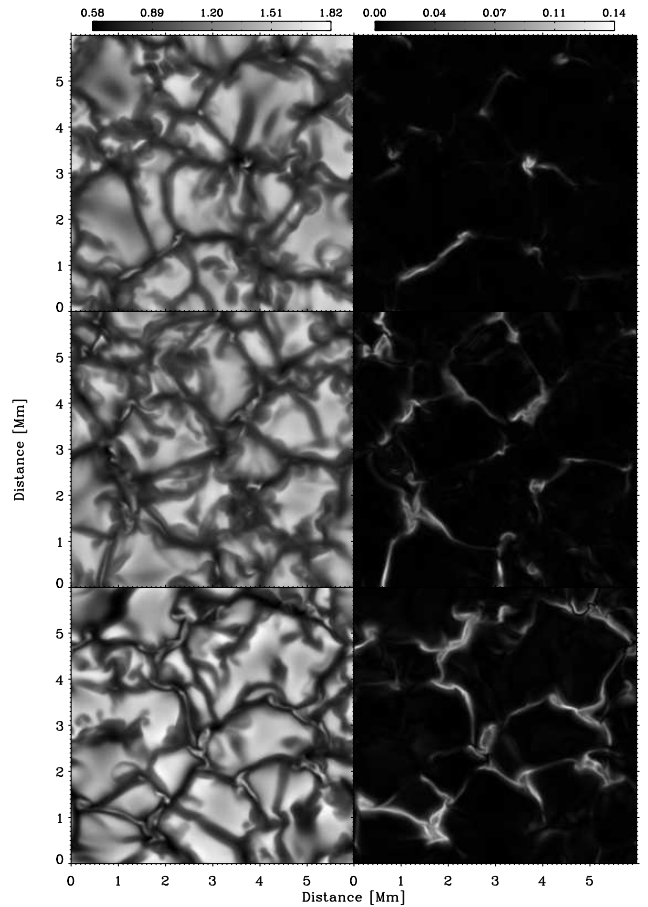}
\caption{Continuum (left panels) and total polarization (right panels) computed
as $\int (Q^2+U^2+V^2) \mathrm{d} \lambda / \int I \mathrm{d}\lambda$ for the three snapshots
considered of 30 G (upper panels), 80 G (middle panels) and 140 G (lower panels) 
of average magnetic field at $\log \tau_{5000}=-1$.}
\label{fig:continuum_maps}
\end{figure}

\subsection{Degradation and phase-diversity reconstruction}
The synthetic profiles at full resolution are degraded to the IMaX spatial
and spectral resolution. IMaX is a filter-polarimetric instrument
based on Fabry-Perot interferometers, so that the spectral point spread function (PSF)
of the double passage is known analytically. This PSF has very extended tails
that would make it necessary to synthesize a large portion of the spectrum to
correctly take into account the smearing. Instead, we follow the approach
of \cite{lagg_imax10} who substituted the real PSF by a Gaussian PSF. It
has been verified empirically that a Gaussian of 85 m\AA\ ($\sim 2.9$ km s$^{-1}$) of full width 
at half-maximum (FWHM) gives a good representation of the effect of the real PSF,
taking into account the contribution of secondary lobes. The
smeared profiles are then sampled at 12 wavelength points to simulate the L12-2
observing mode of IMaX \citep{imax11}, from $-192.5$ m\AA\ to $+192.5$ m\AA\ in 
steps of 35 m\AA.

Concerning the spatial smearing, we have simulated all image degradation effects
that have been considered for Sunrise/IMaX \citep[see][for a detailed analysis]{santiago_vargas09}.
Adding all aberrations, a value of $\lambda/5$ for the rms wavefront error (WFE)
has been measured as the actual performance of the instrument \citep{imax11}. We
consider the same aberration but divided in several contributions. 
The considered telescopic aberrations (up to the 45th Zernike polynomial) 
produce a wavefront with an rms WFE of $\lambda/9$. These aberrations are slowly
varying in time so we consider them fixed during the full polarimetric modulation and
spectral scanning. The value of the coefficients
associated to each Zernike polynomial that conform the wavefront telescopic 
aberration are described by uncorrelated 
Gaussian distributions. The atmosphere is accounted for by considering a wavefront with
turbulent Kolmogorov statistics \citep{noll76,lofdahl_scharmer94}, again with
an rms WFE of $\lambda/9$. While this atmospheric contribution might be considered large
for the case of IMaX/SUNRISE, including it here allows our
results to be valid for ground-based observations with adaptive optics systems. 
This atmospheric contribution might be even considered large for the case of IMaX/SUNRISE. Assuming
the presence of an image stabilization system, we neglect the presence of tip and tilt.
We also consider a ripple WFE of $\lambda/9$ associated
to the polishing that we simulate as a screen with a von Karman 
spectrum with an external scale of 30 cm that simulates the diameter
of the polishing tool. Considering the previous contributions, the total 
rms of the WFE obtained as the addition in quadrature is of the order of $\lambda/5$. Furthermore, we also
take into account the effect of the finite size of the CCD pixels,
which do not sample the image with Dirac delta functions but with 
a small amount of diffraction. Finally, the resulting PSF is then 
convolved with the monochromatic modulated images and noise at the
level of 10$^{-3}$ in units of the continuum intensity is added.
After demodulation, we obtain the degraded monochromatic images of the Stokes parameters.
Additionally, since we apply image reconstruction techniques based on 
the phase-diversity technique, we also simulate images with exactly the
same wavefront but including an additional optimal defocus, in which the
peak-to-valley defocus wavefront equals the observation wavelength \citep{lee_optimaldefocus99}. For the case
of IMaX, this optimal defocus amounts to $\sim 1.81\lambda$.

Our aim is to compare the asymmetries of Stokes $V$ in the degraded
profiles with those found in the original one and in the phase-diversity reconstructed data. 
Since we know exactly the PSFs of the focused and defocused
images, we could apply a standard phase-diversity reconstruction algorithms \citep{gonsalves79,paxman92,lofdahl_scharmer94}
and obtain a perfect reconstruction using:
\begin{equation}
\hat{F} = \frac{\hat{D}_0 \hat{P}_0^\dag + \gamma \hat{D}_k \hat{P}_k^\dag}{|\hat{P}_0|^2+\gamma |\hat{P}_k|^2},
\end{equation}
where $\hat{P}_0$, $\hat{P}_k$, $\hat{D}_0$ and $\hat{D}_k$ are the Fourier transforms of the focused PSF, defocused PSF,
focused image and defocused image, respectively, and $\hat{F}$ is the Fourier transform of the
reconstructed image. In real situations, the PSFs are not known and
the reconstruction algorithm is preceded by an iterative process in which
the PSFs are estimated by maximizing a merit function. In order to overcome this
iterative process but still consider uncertainties in the estimation of the PSFs, we make
the assumption that the projection of the final wavefront on every Zernike polynomial
is known with 10\% relative error. This figure is representative
of extensive simulations carried out with IMaX data \citep[e.g.,][]{santiago_vargas09}. This
error is representing the uncertainties that one would have when using optimization methods 
for the estimation of the wavefront. From the                                                                              
reconstructed focused and defocused PSFs, we obtain the reconstructed image following \cite{gonsalves79,paxman92,lofdahl_scharmer94}.
Because of the presence of noise and the diffraction limit of the telescope, it is 
fundamental to filter out high frequencies from the resulting image \citep{lofdahl_scharmer94}. These
high frequencies cannot be recovered. Additionally,
the noise on the reconstructed images present a Gaussian distribution with a variance
larger than that of the original focused and defocused images. The main problem
is that this noise is spatially correlated \citep{meynadier99}. This leads to an aesthetically
unappealing web-like noise which avoids using many profiles in the quiet-Sun because
their amplitudes are already very low. Furthermore, this 
noise introduces artifacts that perturb the appearance of the Stokes profiles. This
effect is also discussed by \cite{klaus11} with different deconvolution methods.

\begin{table*}
\caption{Statistical information about profiles}
\label{tab:classification_profiles}
\centering
\renewcommand{\footnoterule}{}  
\begin{tabular}{c|c|ccccccc}
\hline \hline
Snapshot & Case & Above & 1 lobe & 2 lobes & 3 lobes & 4 lobes & Amplitude & Area \\
 & & threshold$^{\mathrm{a}}$ &  &  &  &  & assymetry$^{\mathrm{b}}$ & asymmetry$^{\mathrm{c}}$  \\
\hline
30 G & Full (IMaX) & 36\% & 14.61\% & 80.36\% & 4.32\% & 0.05\% & $8.65 \pm 38.6$ & $6.68 \pm 42.4$\\
- & Degraded (IMaX) & 11\% & 0.75\% & 99.25\% & 0.01\% & 0.01\% & $8.65 \pm 38.6$ & $-5.72 \pm 15.8$ \\
- & Reconstructed (IMaX) & 5\% & 17.44\% & 82.48\% & 0.10\% & 0.03\% & $1.79 \pm 14.7$ & $15.58 \pm 34.92$\\
- & Full (all $\lambda$) & 54\% & 25.12\% & 64.78\% & 9.88\% & 0.07\% & $28.10 \pm 33.9$ & $3.02 \pm 29.2$\\
\hline
80 G & Full (IMaX) & 68\% & 14.00\% & 80.90\% & 4.90\% & 0.04\% & $9.04 \pm 39.2$ & $8.12 \pm 42.5$\\
- & Degraded (IMaX) & 34\% & 1.27\% & 98.71\% & 0.02\% & 0.01\% & $4.53 \pm 16.2$ & $-0.69 \pm 18.0$\\
- & Reconstructed (IMaX) & 21\% & 14.60\% & 84.74\% & 0.65\% & 0.01\% & $22.10 \pm 31.8$ & $16.54 \pm 32.6$\\
- & Full (all $\lambda$) & 85\% & 16.26\% & 69.55\% & 14.07\% & 0.06\% & $27.67 \pm 37.6$ & $5.00 \pm 30.6$\\
\hline
140 G & Full (IMaX) & 85\% & 12.61\% & 84.30\% & 3.05\% & 0.01\% & $18.07 \pm 40.0$ & $19.12 \pm 43.5$\\
- & Degraded (IMaX) & 71\% & 2.42\% & 97.32\% & 0.26\% & 0.00\% & $13.53 \pm 20.6$ & $11.46 \pm 23.4$ \\
- & Reconstructed (IMaX) & 34\% & 13.19 & 85.49\% & 1.31\% & 0.00\% & $21.66 \pm 32.3$ & $20.18 \pm 34.2$\\
- & Full (all $\lambda$) & 95\% & 7.04\% & 80.55\% & 12.34\% & 0.05\% & $28.87 \pm 38.4$ & $13.50 \pm 32.2$\\
\hline
\end{tabular}
\begin{list}{}{}
\item[$^{\mathrm{a}}$] Percentage of profiles with Stokes $V$ signal above the noise threshold.
\item[$^{\mathrm{b}}$] Mean and standard deviation of amplitude asymmetry.
\item[$^{\mathrm{c}}$] Mean and standard deviation of area asymmetry.
\end{list}
\end{table*}


\section{Circular polarization asymmetries}
Circular polarization asymmetries are usually defined in terms of the relative
area ($\delta A$) and amplitude ($\delta a$) imbalances between the blue and the red lobes of Stokes $V$
profiles \citep[e.g.,][]{solanki_stenflo86}:
\begin{eqnarray}
\delta a &=& \frac{a_b-a_r}{a_b+a_r} \\
\delta A &=& s \frac{\int{V(\lambda) d \lambda }}{\int{|V(\lambda)| d \lambda }},
\end{eqnarray}
where $a_r$ and $a_b$ refer to the amplitude of the red and blue lobes, respectively.
The factor $s$ gives the sign of the area asymmetry, and it is chosen equal to the
sign of the bluest peak of Stokes $V$, following \citep{martinezpillet97}. 

\begin{figure*}[!t]
\centering
\includegraphics[width=0.59\textwidth,angle=90]{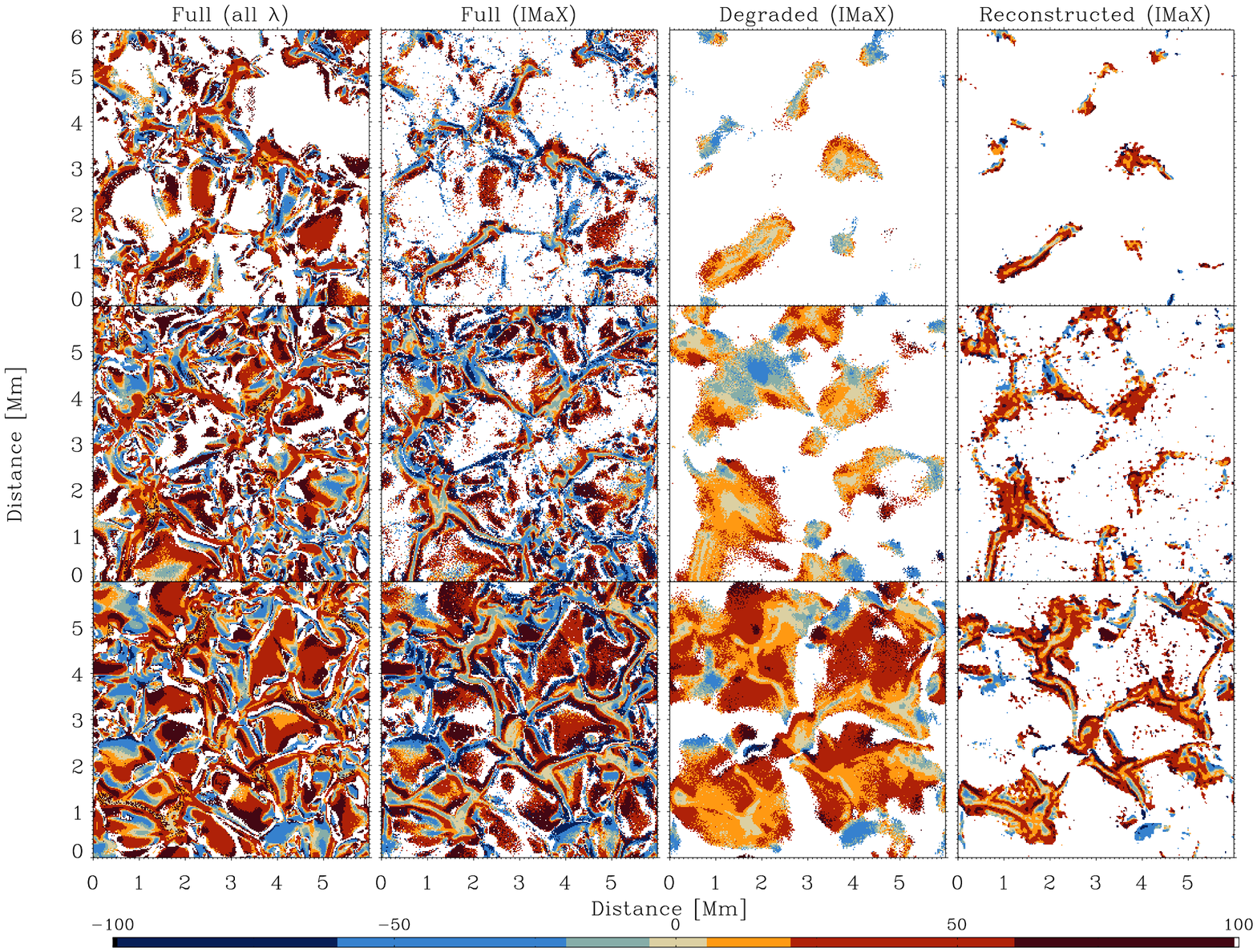}
\includegraphics[width=0.59\textwidth,angle=90]{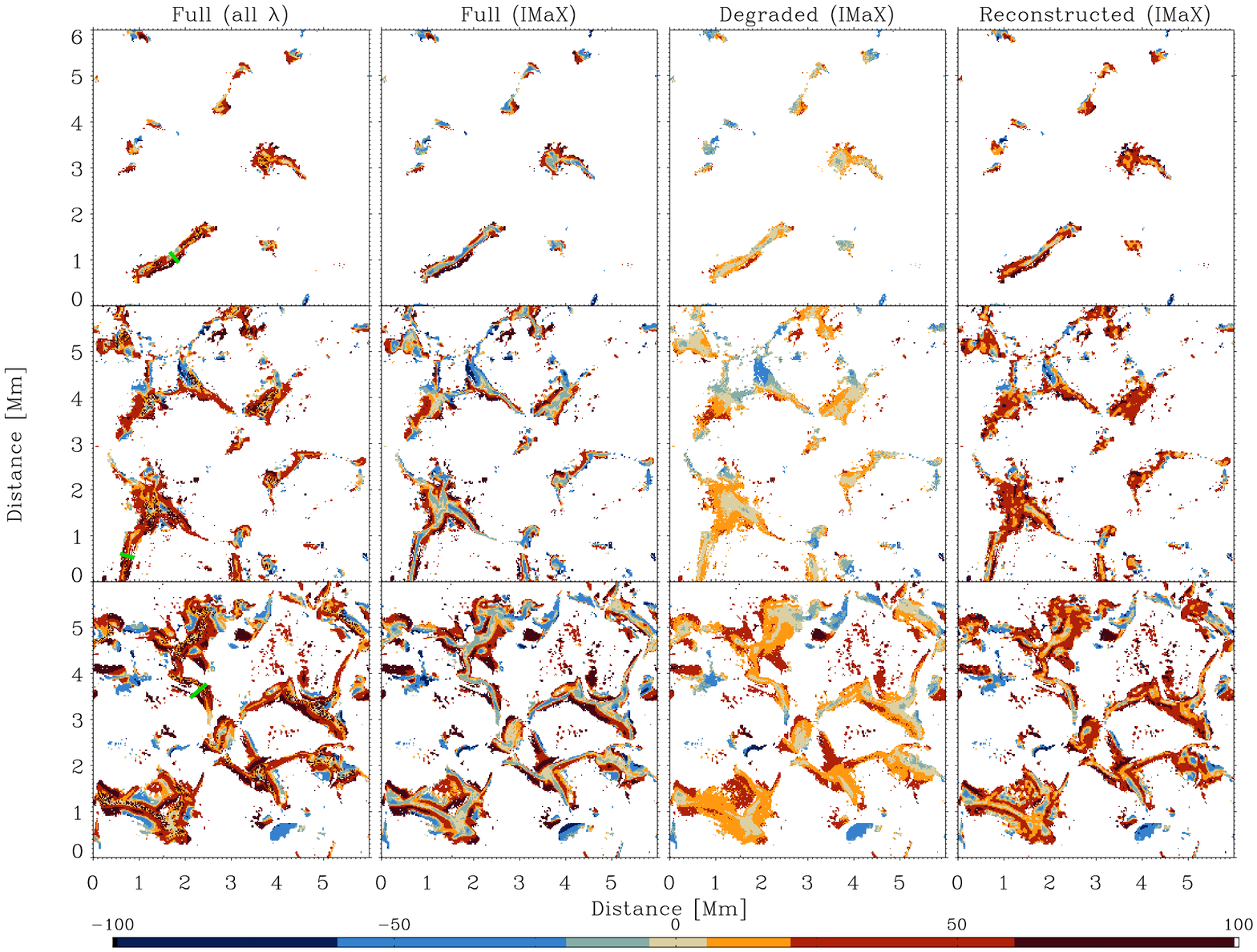}
\caption{Amplitude asymmetry maps for the four considered smearing cases (columns) and
for the three values of the seed magnetic field used (rows): 30 G, 80 G and 140 G. The upper panels show
the map locations where the Stokes $V$ profiles are above the threshold in each individual case.
The lower panels display only locations where the Stokes $V$ profiles are above the
threshold simultaneously in the four considered cases. The small green lines at positions
$(2,1)$ for the upper panels, $(0.5,0.5)$ for the middle panels and $(2.3,3.8)$ for the lower panels show the location
of the cuts along which the Stokes $V$ profiles of Fig. \ref{fig:stokes_cuts} have been obtained.}
\label{fig:amplitude_asymmetry_maps}
\end{figure*}

\begin{figure*}[!t]
\centering
\includegraphics[width=0.59\textwidth,angle=90]{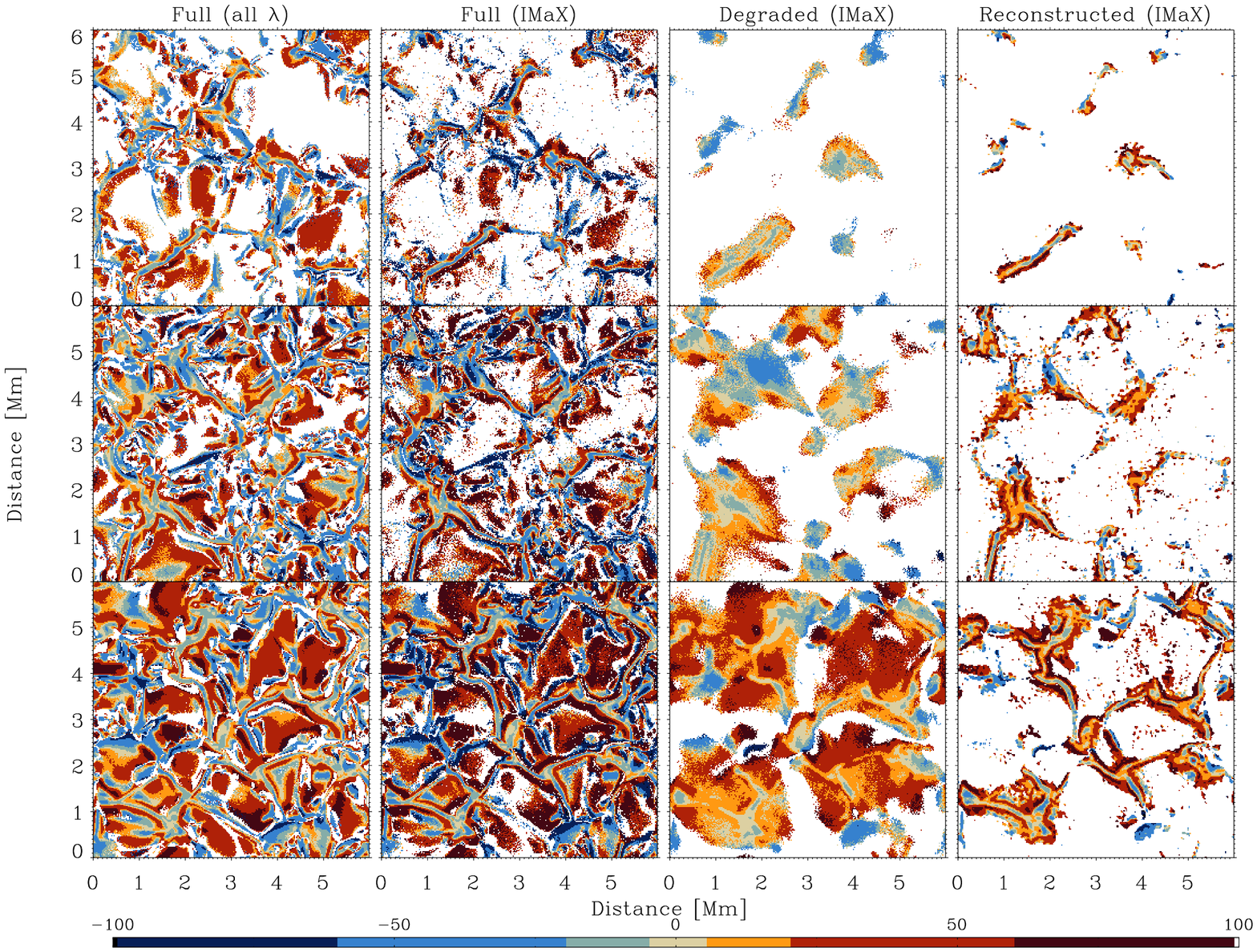}
\includegraphics[width=0.59\textwidth,angle=90]{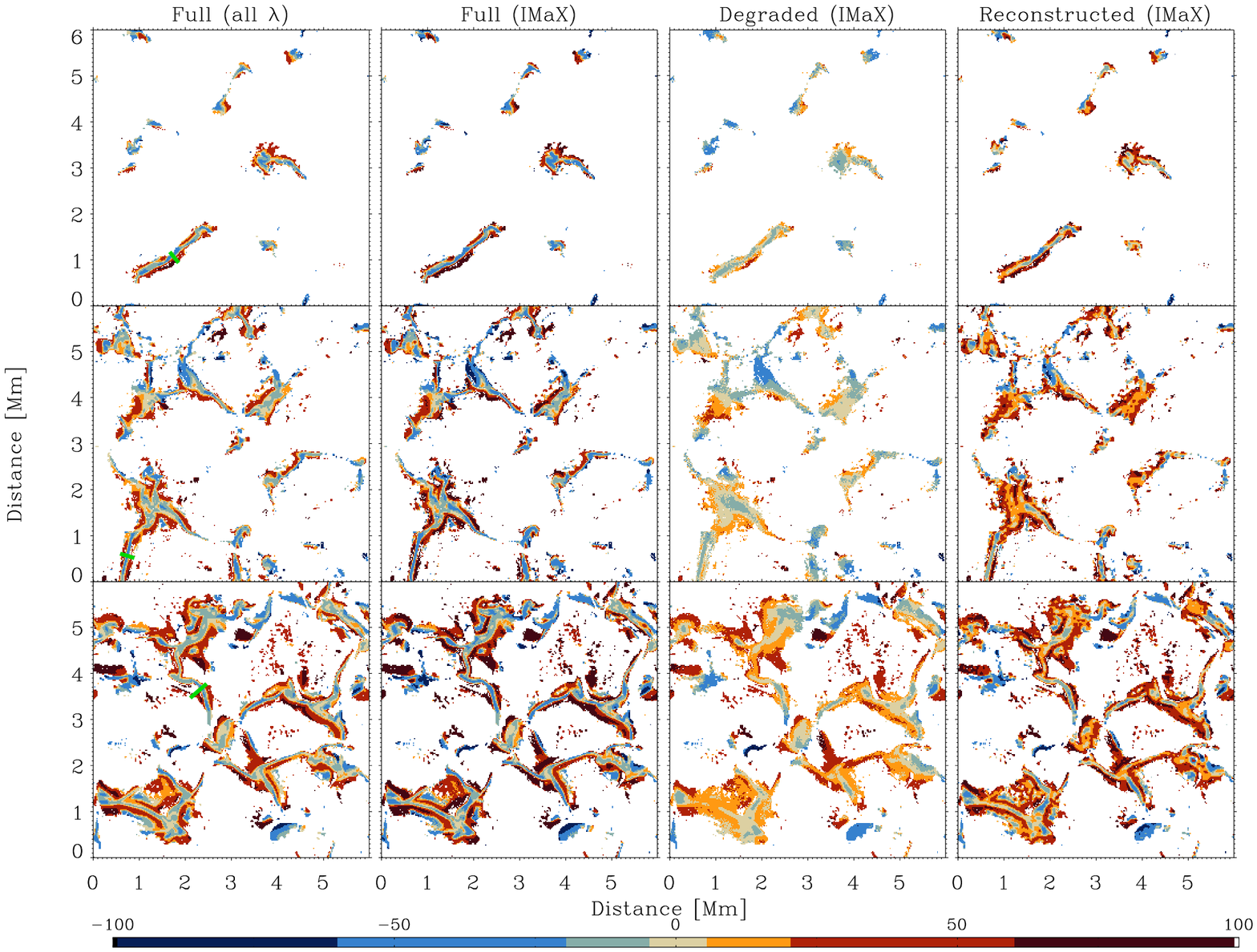}
\caption{Similar to Fig. \ref{fig:amplitude_asymmetry_maps} but for the area asymmetry.}
\label{fig:area_asymmetry_maps}
\end{figure*}

\begin{figure*}[!t]
\centering
\includegraphics[width=0.7\textwidth,angle=90]{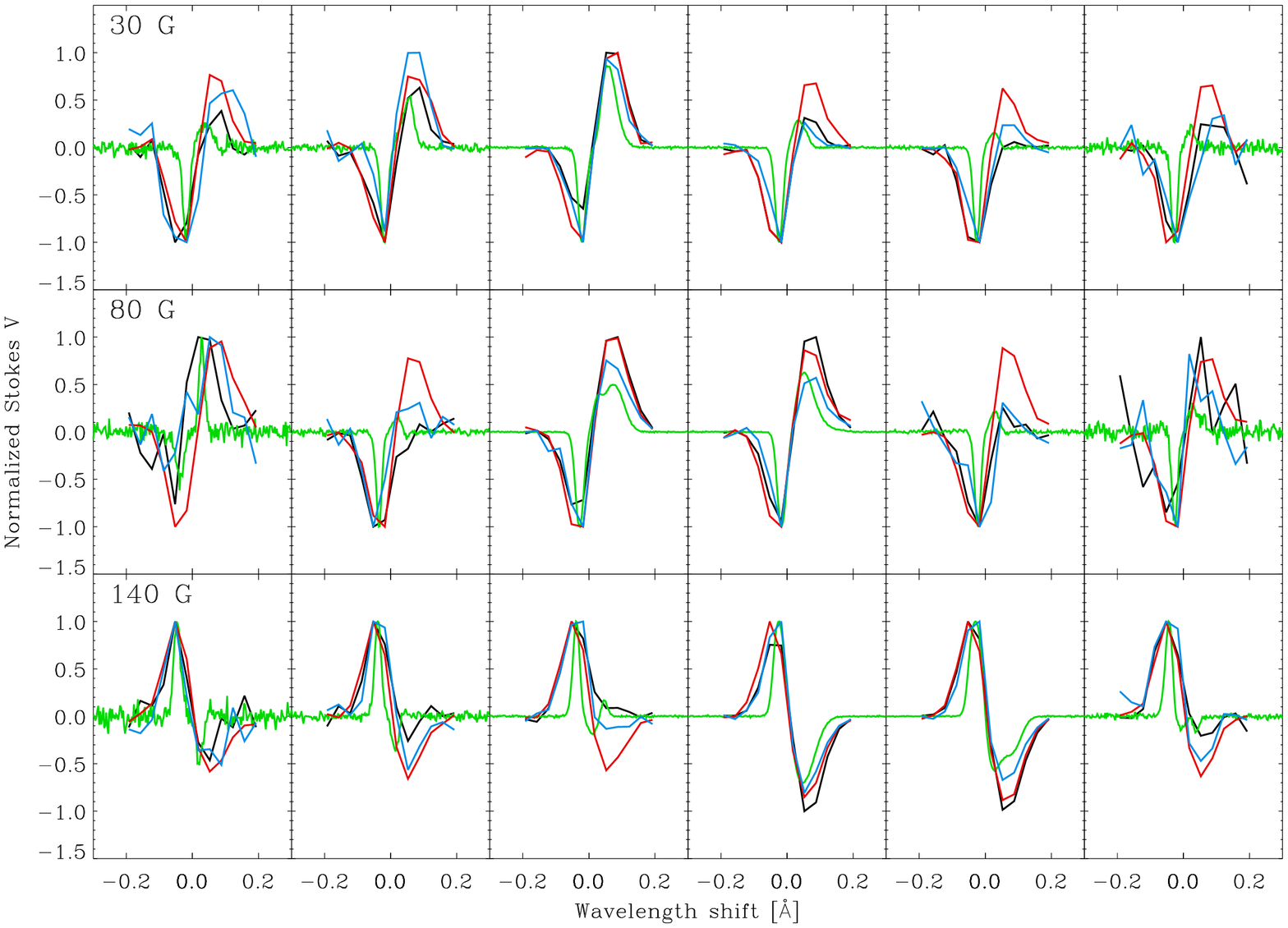}
\caption{Stokes $V$ profiles normalized to the peak amplitude across the cuts shown in Fig. \ref{fig:amplitude_asymmetry_maps}
and \ref{fig:area_asymmetry_maps} in green. Black profiles correspond to the ``Full (IMaX)'' case, while green 
ones are associated with the ``Full (all $\lambda$)'' case.
Red profiles correspond to the ``Degraded (IMaX)'' case and blue profiles are the ``Reconstructed (IMaX)'' ones.}
\label{fig:stokes_cuts}
\end{figure*}

These definitions of asymmetries can only be strictly applied to standard two-lobed
antisymmetric profiles (one-lobed profiles can also be considered as a
degenerate two-lobed profile). However, simulated Stokes $V$ profiles often show a panoply
of profiles whose spectral shapes are completely different to the standard
antisymmetric profile. Table \ref{tab:classification_profiles} displays the
percentage of Stokes $V$ profiles with different number of lobes
on the three snapshots considered. Although the
majority of profiles are standard two-lobed profiles, there are many
locations where 1-lobed or 3-lobed profiles are found. This situation is more
critical for the case in which the full spatial and spectral resolution is considered. 
To overcome this problem, we have restricted our
analysis to one-lobed and two-lobed profiles, leaving for the future the
more complex analysis of profiles with more lobes. In such a case, we think that
it will be necessary to update the definition of asymmetries. Interestingly, recent 
high spatial resolution observations (Mart\'{\i}nez Gonz\'alez et al. 2011, in preparation) also show the appearance
of an increasingly larger fraction of non-standard profiles.

In order to analyze asymmetries, we consider four different cases that
will help us understand how Stokes $V$ profiles are distorted by the
presence of spatial and spectral smearing. The first case, which we label as ``Full (all $\lambda$)'' 
consists of the line profiles at full spatial and spectral resolution. The next, 
labeled as ``Full (IMaX)'', are the profiles at full spatial resolution but 
degraded to the IMaX L12-2 observing mode spectral resolution. We also consider
two observationally realistic cases. The first one is the fully degraded, labeled as
``Degraded (IMaX)''. The standard deviation of the noise added a-posteriori in units of the continuum intensity
in these snapshots is $10^{-3}$. The last one is the phase-diversity reconstructed
case, labeled ``Reconstructed (IMaX)''. The standard deviation of the noise
increases to $3-4 \times 10^{-3}$ in units of the continuum intensity after the reconstruction.

\begin{figure*}
\centering
\includegraphics[width=0.49\textwidth]{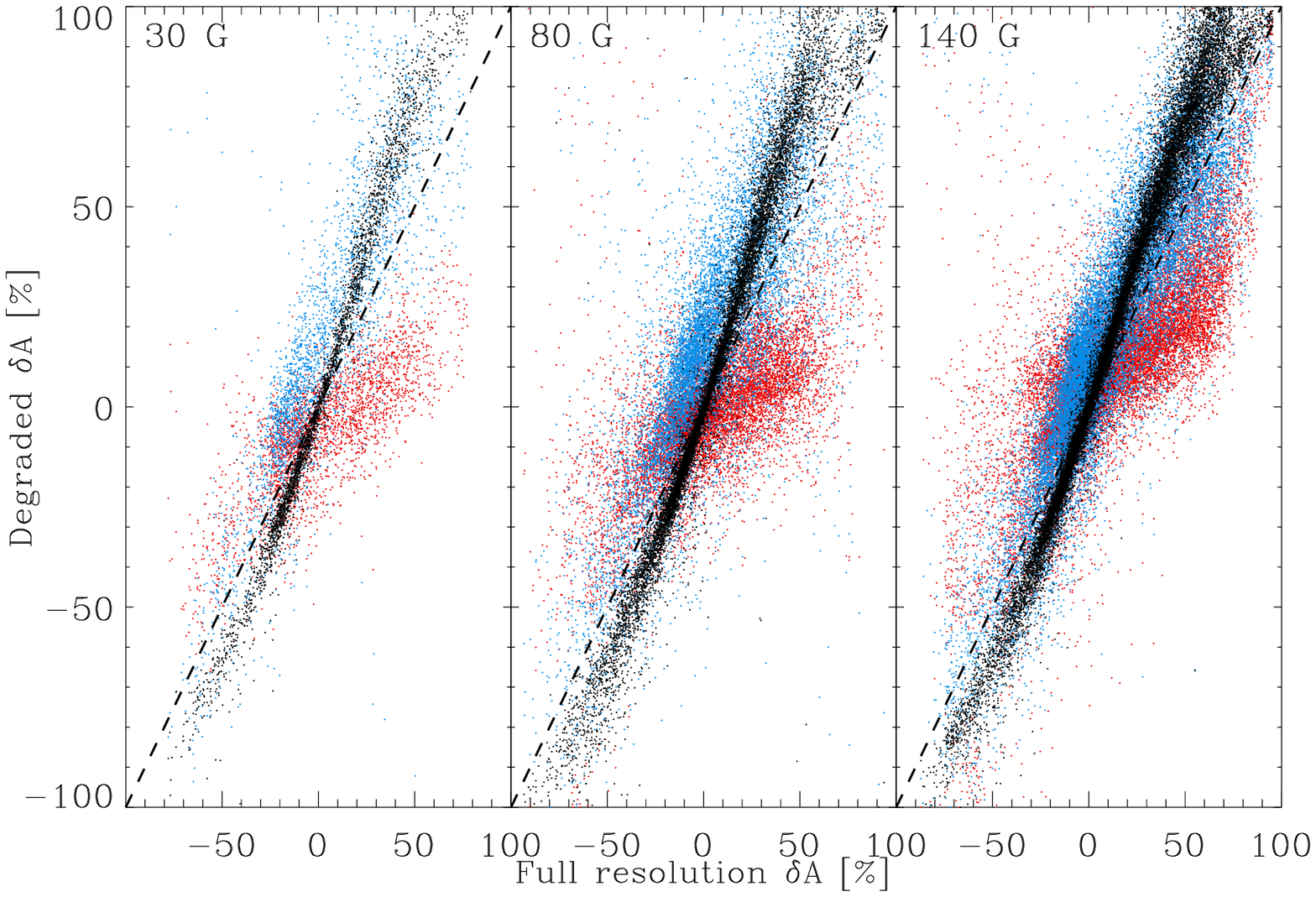}
\includegraphics[width=0.49\textwidth]{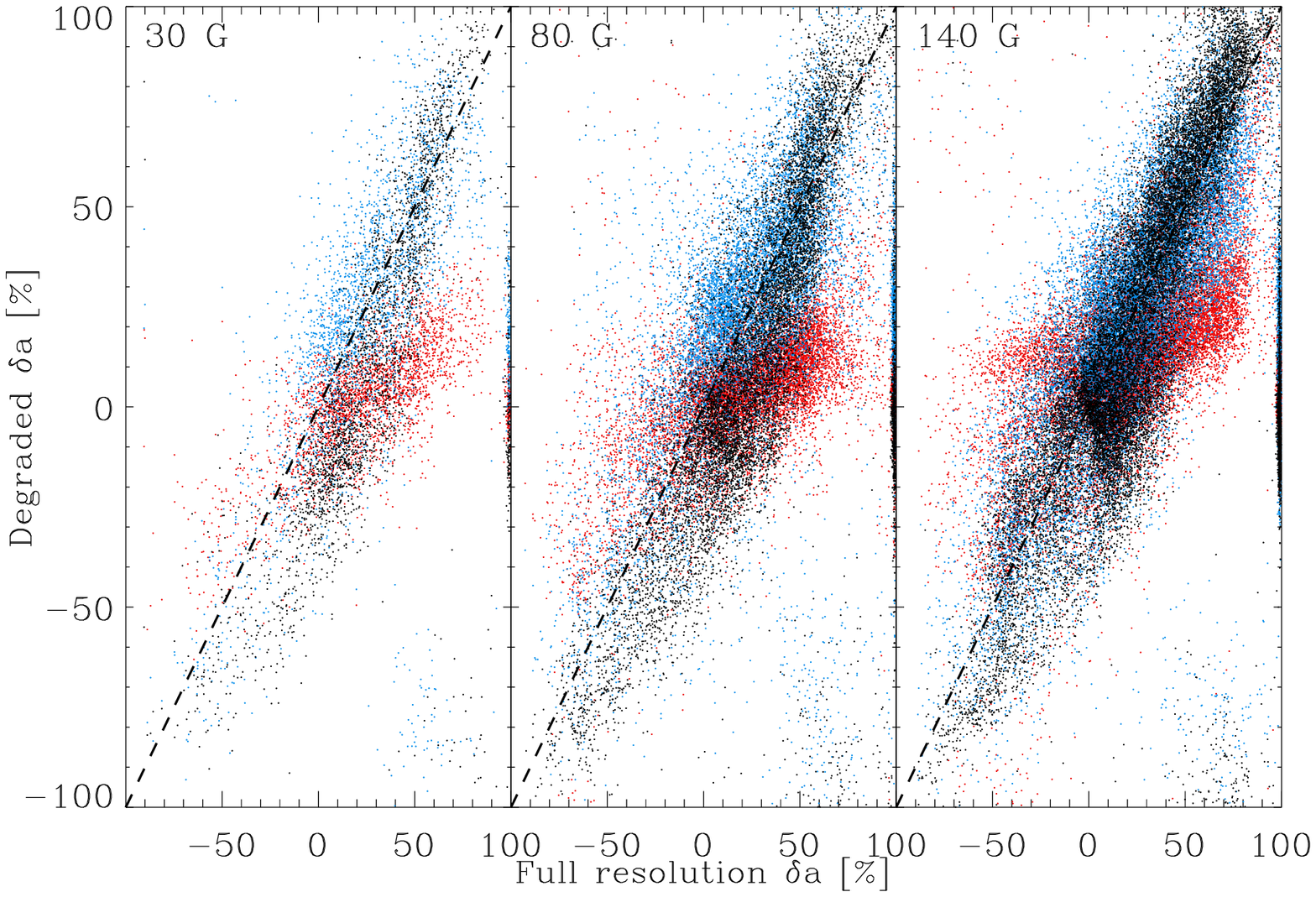}
\includegraphics[width=0.49\textwidth]{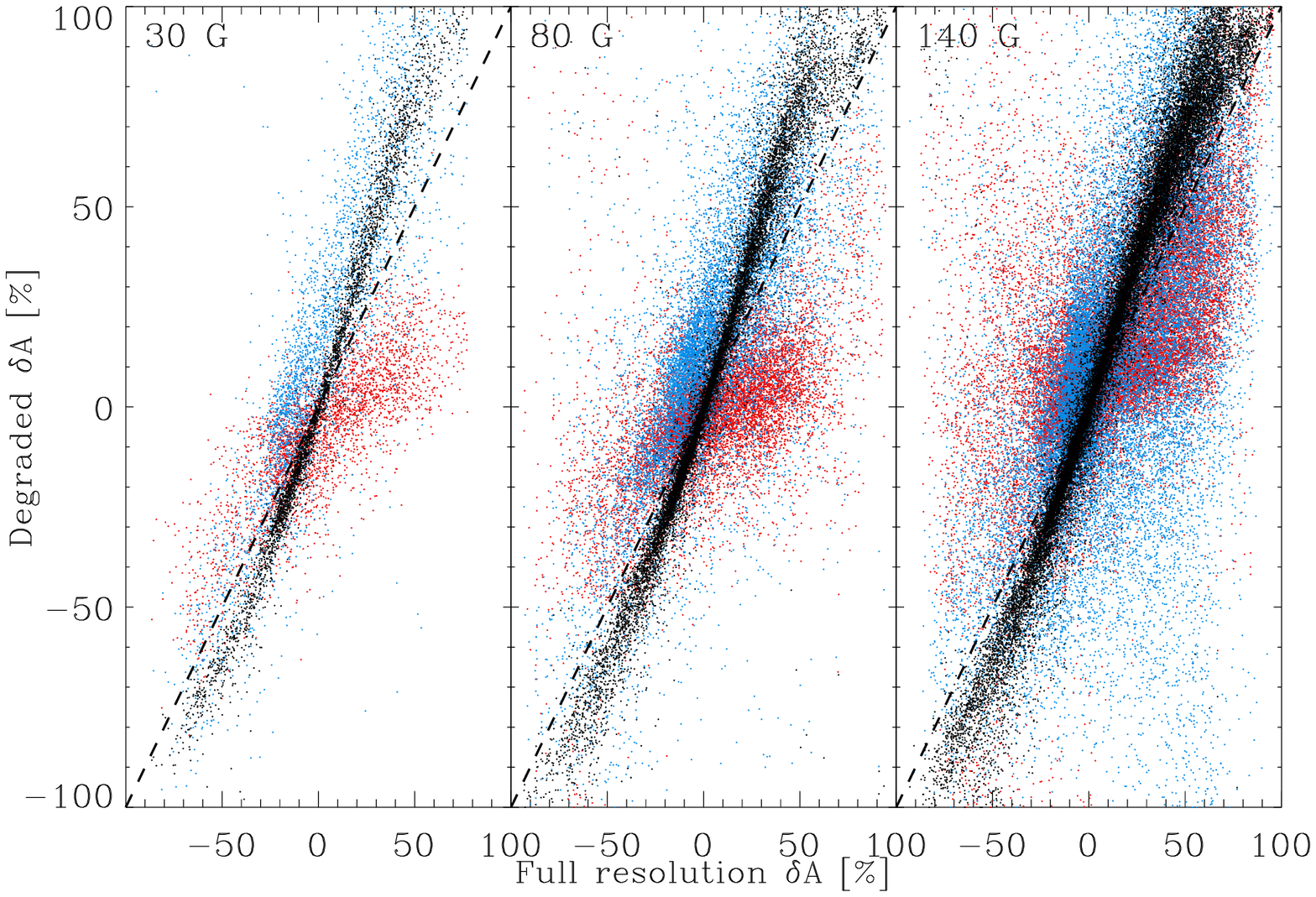}
\includegraphics[width=0.49\textwidth]{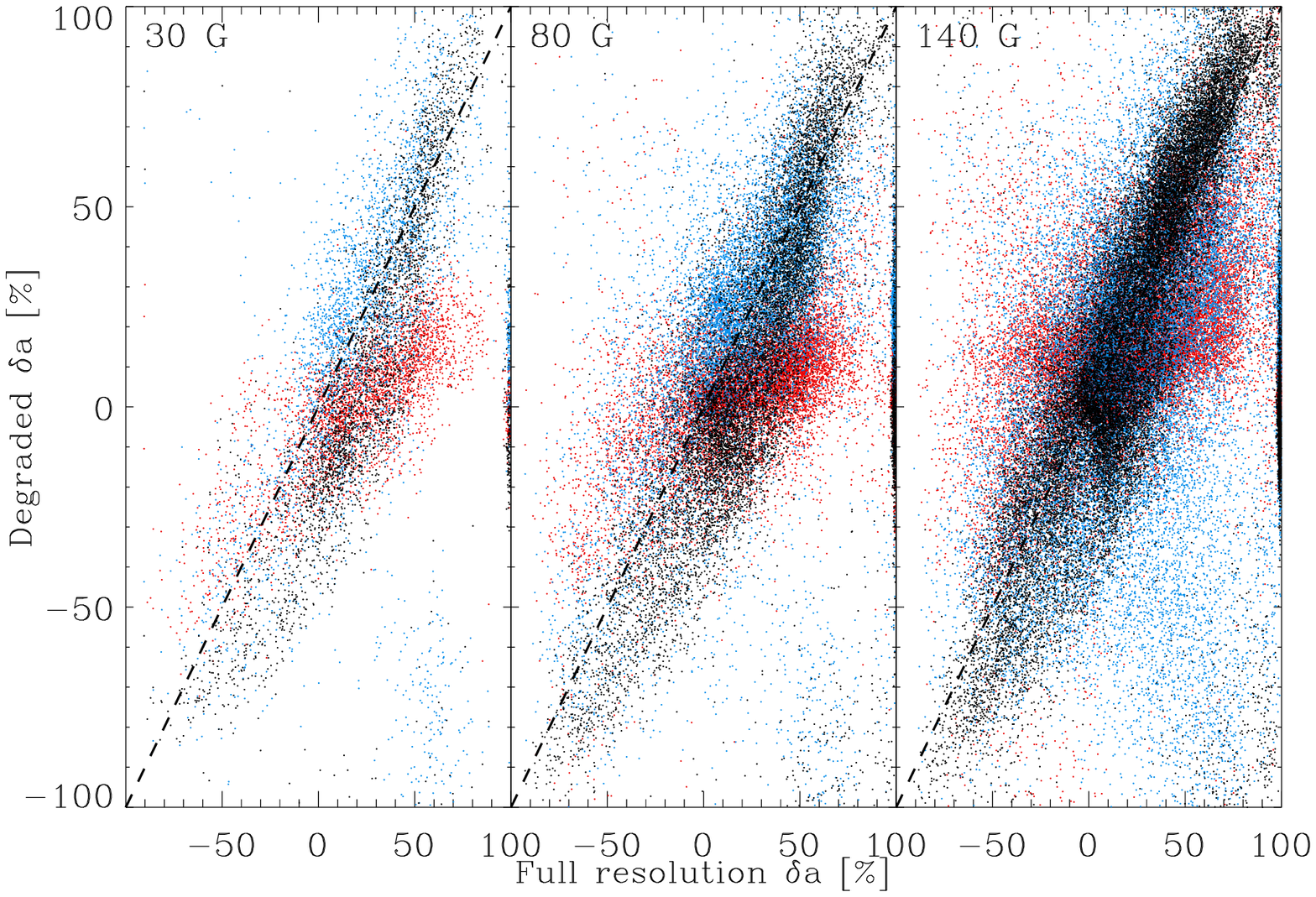}
\caption{Left panels: area asymmetries obtained on the degraded snapshots (black for ``Full (IMaX)'' case,
red for ``Degraded (IMaX)'' and blue for ``Reconstructed (IMaX)'') versus the original ``Full (all $\lambda$)'' 
area asymmetries. Right panels: same but for amplitude asymmetry. The upper panels show results when the
wavefront is known with 10\% uncertainty, while the lower panels display what happens when the
uncertainty increases to 25\%.}
\label{fig:asymmetries_full_vs_degraded}
\end{figure*}

\subsection{Spatial variation}
Figures \ref{fig:amplitude_asymmetry_maps} and \ref{fig:area_asymmetry_maps} show the amplitude
and area asymmetries for all considered cases, respectively. The upper panels show all
map locations where the Stokes $V$ profiles are above the threshold for each individual
case. For the sake of a better visualization, the bottom panels display only map locations
whose Stokes $V$ profiles are above 3 times the value of the standard deviation
of the noise in all cases simultaneously. In this case, the ``Reconstructed (IMaX)'' case
is the one controlling such map locations because of the enhanced noise level.

From left to right
we show what happens when the spatial/spectral resolution is degraded, while the last panel displays
the phase-diversity reconstructed case. The three cases considered (each row) 
correspond to the three snapshots with different seed magnetic fields, where only
profiles above the threshold of 3 times the noise level have been considered. Obviously, 
the number of selected profiles increases as the seed field increases because of the 
enhanced Stokes $V$ amplitudes. 

Concerning the high-resolution maps, we find predominantly positive area asymmetries in
granules, similar to the previous analysis of \cite{khomenko_shelyag05} and \cite{shelyag07} for the
Fe \textsc{i} lines at 630 nm and 1.56 $\mu$m. This
result is compatible with the predominance of positive area asymmetries in granules
found in observations \citep[e.g.,][]{socas_pillet_lites04}. This predominance
is also found for amplitude asymmetries. Negative asymmetries are found in intergranular
lanes and in strong kG acummulations. This result found in the high-resolution maps
is essentially incompatible with the observational 
results of \cite{socas_pillet_lites04}, who find positive area and amplitude asymmetries in lanes
and the network. We show below that both results can be reconciled after image degradation.

Asymmetries in the high-resolution maps can be understood under the flux tube model with a canopy developing
with height \citep{bellotrubio00}. The inner parts of the flux tubes coincide with intergranular
lanes in the simulations and present strongly negative area asymmetries with not so strongly
negative amplitude asymmetries. They are generated by strong downflows associated with a field
strength that decreases with height \citep{shelyag07}. The granule-intergranule transition with positive
area and amplitude asymmetries can be explained by rays crossing the canopy of the flux tube.

Apparently, the spectral degradation of the data at
full spatial resolution has a relatively small impact on the spatial distribution
of asymmetries, as seen from a comparison of the first and second columns. The 
filamentary structure of the asymmetries associated with intergranular lanes is 
maintained, while the profiles observed in granular regions are lost because the spectral
degradation pushes them below the noise threshold. However, we can see that the
spectrally degraded data presents larger (in absolute value) area and amplitude asymmetries than the
original dataset, being this difference especially relevant on the cores of
intergranular lanes. This can be easily understood because spectral degradation tends
to increase relative area and amplitude asymmetries \cite{solanki_stenflo86}, although the
dependency on the specific shape of the Stokes $V$ profile is crucial.

Spatial degradation changes completely the scenario. The main reason is that spatial
smearing produces two effects: i) profiles with very different asymmetries are mixed, and ii)
a decrease on the Stokes $V$ amplitude, thus pushing the signal
in many pixels below the noise threshold. Still, the pixels associated with the
strongest signals remain (11\%, 38\% and 70\% of the pixels for the 30 G, 80 G
and 140 G snapshots, respectively) and the filamentary structure is somehow
lost and transformed into a blobby structure. Typically, asymmetries tend to be organized inside these blobs
so that zero or very small asymmetries are found in the cores while they increase
in absolute value towards the borders, where profiles are lost below the noise
threshold. This behavior is not exactly the same of what happens at larger spatial resolutions,
where the cores of intergranular lanes present negative asymmetries while they become 
positive on the granule-intergranule transition region. From a 
general perspective, we can say that spatial smearing makes the negative asymmetry of narrow 
intergranular lanes appear with smaller absolute values while broadening the
region of positive asymmetry.

The area and amplitude asymmetries found in the spatially degraded snapshots tend to be more
compatible with observations. On average, asymmetries tend to be positive, with
a larger absolute value for amplitude asymmetries than for area asymmetries. Stokes $V$ 
profiles in integranular lanes have now negative area asymmetries not larger than 
15-20\% or even compatible with zero. The only exceptions are isolated patches
with strong negative asymmetries.

It is evident from Figs. \ref{fig:amplitude_asymmetry_maps} and \ref{fig:area_asymmetry_maps}
that area asymmetries are slightly less affected by the spatial and spectral smearing than amplitude
asymmetries. It is specially relevant in the central parts of internetwork
lanes, where strong negative amplitude asymmetries show up when reducing the spectral resolution
and these are transformed into almost amplitude symmetric profiles when degrading 
the spatial resolution. On the contrary, relevant negative area asymmetries are still present
on internetwork lanes in the spatially and spectrally smeared data.


The main point of this work is to verify to what extend asymmetries are recovered
after image reconstruction techniques. The rightmost panels of Figs. \ref{fig:amplitude_asymmetry_maps} 
and \ref{fig:area_asymmetry_maps} demonstrate that an improvement on the spatial
location of asymmetries occurs. Obviously, the enhanced noise level and its
spatially correlated character produces that the number of pixels with detected
signals decreases (6\%, 22\% and 43\% of the pixels for the 30 G, 80 G
and 140 G snapshots, respectively). Comparing with Fig. \ref{fig:continuum_maps}, only pixels associated to strong 
magnetic concentrations can be correctly analyzed. Consequently, one should be careful when 
analyzing reconstructed data to take into account this bias. Several conclusions
can be extracted from the reconstructed images. First, 
the structures are much more compact than in the degraded case, a consequence of 
the efficient reduction of aberrations performed by the phase-diversity
algorithm, even with 10\% relative errors on
the projections of the wavefront on the Zernike polynomials. Second, the large
gradient of asymmetries from the core of intergranular lanes to the surroundings is partially
restored. However, the large negative
value of $\delta A$ and $\delta a$ found in the very central cores are not fully recovered.
Generally, in these regions, image reconstruction cannot generate strong negative
asymmetries if they were not present in the degraded images. Likewise, if they
were present in the degraded maps, they will be enhanced in the image reconstruction
process.

\subsection{Line profiles}
Synthetic Stokes $V$ profiles for the three snapshots considered are shown 
in Fig. \ref{fig:stokes_cuts} for the cuts indicated in 
Figs. \ref{fig:amplitude_asymmetry_maps} and \ref{fig:area_asymmetry_maps} in green.
The Stokes profiles are normalized to peak amplitude so one has to take into account
that spatially and/or spectrally averaged signals possess a smaller amplitude. The profiles 
of the full spectral and spatial resolution are shown in green. The profiles of the
full spatial resolution but spectral resolution degraded to IMaX are shown in black. The
fully degraded profiles are shown in red and the phase-diversity recovered ones are
plotted in blue. 

The first impression is that the full spatial and spectral resolution profiles
are very narrow and the presence of several components (several lobes on the
red lobe) along the LOS can be easily witnessed. When the spectral
resolution is degraded to 85 m\AA, the profiles are broadened, with strong modifications
of the relative amplitudes of the two lobes. As a byproduct, the full spatially/spectrally 
degraded profiles tend to be much more symmetric and only in those cases in which the full resolution
profile is extremely asymmetric, some asymmetry remains on the degraded profiles.

It is interesting to note that the phase-diversity reconstructed profiles tend
towards the full spatial resolution case with IMaX spectral resolution, as expected.
This means that an image reconstruction algorithm is able to partially extract the
information necessary to recover the Stokes $V$ profiles. This is especially
relevant for the very asymmetric profiles.

\subsection{Statistical properties}
We have tested that image reconstruction based on phase-diversity does a 
good job at cancelling the spatial smearing of polarimetric signals. Furthermore,
although many points are lost below the enhanced noise level, the spatial appearance
of asymmetries is not strongly modified. However, it is important to verify other properties.
We focus now on the statistical properties of asymmetries. 

Fig. \ref{fig:asymmetries_full_vs_degraded} displays the area (left panels)
and amplitude (right panels) asymmetries computed on the degraded snapshots versus 
those obtained on the full resolution snapshots. The upper panels correspond
to the case in which the wavefront is known with 10\% uncertainty, whereas the
lower panels are associated to the case in which the uncertainty increases to
25\%. Red points correspond to the
``Degraded (IMaX)'' case. Black points correspond to the ``Full (IMaX)'' case,
while the blue points are associated to the ``Reconstructed (IMaX)'' case. 
Since the number of points with profiles above the noise level is different
for each case, we consider only those points which produce signal above the threshold in the
four cases simultaneously. Essentially, this leaves only those points on the phase-diversity reconstructed
images, as shown in the lower panels of Figs. \ref{fig:amplitude_asymmetry_maps} and 
\ref{fig:area_asymmetry_maps}. 

The spectral degradation produced by IMaX induces that area asymmetries (in
absolute value) are systematically overestimated \cite{solanki_stenflo86} because
the black points are above (for positive asymmetries) and below (for negative asymmetries)
the main diagonal.
Concerning the amplitude asymmetry at IMaX spectral resolution, the lack of spectral resolution
induces a spread around the correct value. When the spatial resolution is degraded to 
IMaX resolution, we find that both area and amplitude asymmetries are always
underestimated with respect to the original one. However, while positive asymmetries
are underestimated, negative asymmetries are severely lost. The reason is that these
points with negative asymmetries are associated to profiles with low Stokes $V$ amplitude
and are quickly lost below the noise level \citep[see][for examples on simulations]{khomenko_shelyag05,shelyag07}.

The phase-diversity reconstruction leads to a correction of the area and amplitude
asymmetries such that the overlap with the black points is enhanced. This overlapping
is evident for the case in which the wavefront is known with 10\% accuracy but less
evident when the uncertainty raises to 25\%. As a byproduct, some portion of the
large negative asymmetries appearing on the central parts of intergranular lanes (specifically,
those surrounded by large positive asymmetries) are recovered. On the contrary, isolated strong
negative asymmetries (like that found at position (2.5,3) on the upper panel of Figs. \ref{fig:amplitude_asymmetry_maps}
and \ref{fig:area_asymmetry_maps}) are lost below the noise level.
Consequently, we conclude that the image reconstruction algorithm is able to partially
recover the main properties of area and amplitude asymmetries.

Interestingly, area asymmetries can be 
corrected to be close to the original ones if they are divided by a factor between $0.3$ and
$0.5$. This figure is slightly dependent on the mean magnetic field of the simulation. A 
correction factor of $\sim 0.4$ is a good compromise, although the spatial
smearing is still present and strong asymmetries are fully lost. 


\section{Conclusions}
We have analyzed the impact of phase-diversity image reconstruction algorithms on the 
Stokes $V$ asymmetries. To this end, we have employed three snapshots of 3D
MHD models of solar magneto-convection with different mean magnetic fields, as representative
of regions with different magnetization on the solar surface. We have
synthesized the Fe \textsc{i} line at 5250.2089 \AA\, introduced spatial and
spectral smearing to simulate observations with IMaX, and applied a
phase-diversity reconstruction assuming imperfect knowledge of the wavefront.
We conclude that asymmetries are modified by the smearing process and
that this effect can be alleviated using the reconstruction algorithm. 
Area and amplitude asymmetries tend to be overestimated after the spectral smearing. This is compensated by the
spatial smearing, which tends to underestimate all asymmetries. 
Large negative asymmetries are lost below the noise level because they are
usually associated to low-amplitude Stokes $V$ profiles. Reconstruction
successfully reduces the underestimation of asymmetries and partially
recovers the main characteristics. However, patches of isolated large negative
asymmetries cannot be recovered.
We point out that a large fraction of the pixels cannot be analyzed in the
reconstructed data because of the enhanced noise. This noise
has spatial correlation, which introduces additional problems.
The individual line profiles tend to look more similar to those of the full spatial resolution
case with degraded spectral resolution.

\begin{acknowledgements}
We are grateful to C. Beck for helpful comments. We also thank M. Sch\"ussler and A. V\"ogler for providing the 3D
snapshots used in this work. Financial support by the 
Spanish Ministry of Science and Innovation through projects AYA2010-18029 (Solar Magnetism and Astrophysical 
Spectropolarimetry) and Consolider-Ingenio 2010 CSD2009-00038 is gratefully acknowledged.
\end{acknowledgements}


\end{document}